# A Service Broker Model for Cloud based Render Farm Selection

Ruby Annette
Research Scholar
B.S. Abdur Rahman University

Aisha Banu.W, PhD
Professor, CSE Department,
B.S. Abdur Rahman University

## ABSTRACT
Cloud computing is gaining popularity in the 3D Animation industry for rendering the 3D images. Rendering is an inevitable task in creating the 3d animated scenes. It is a process where the scene files to be animated is read and converted into 3D photorealistic images automatically. Since it is a computationally intensive task, this process consumes the majority of the time taken for 3D images production. As the scene files could be processed in parallel, clusters of computers called render farms can be used to speed up the rendering process. The advantage of using Cloud based render farms is that it is scalable and can be availed on demand. One of the important challenges faced by the 3D studios is the comparison and selection of the cloud based render farm service provider who could satisfy their functional and the non functional Quality of Service (QoS) requirements. In this paper we propose, a frame work for Cloud Service Broker (CSB) responsible for the selection and provision of the cloud based render farm. The Cloud Service Broker matches the functional and the non functional Quality of Service requirements (QoS) of the user with the service offerings of the render farm service providers and helps the user in selecting the right service provider using an aggregate utility function. The CSB also facilitates the process of Service Level Agreement (SLA) negotiation and monitoring by the third party monitoring services.

## 1. INTRODUCTION
Cloud computing is a widely researched topic and many researches are focused on improving the technology and facilitating the use of the technology. One of the concepts that have evolved to felicitate the use of the cloud technology is the Cloud Service Broker (CSB). According to Gartner, the role of a CSB is inevitable to make the user from various other domains embrace the cloud computing technology. The CSB would play a major role as the user from the other domains lack the knowledge of the technical aspects in cloud computing and fear the use of the unknown technology. A CSB would be able to assist the user with the expertise knowledge in the cloud computing technology and assist the end user in indentifying the services and using them without fear or confusion. In the words of Gartner, the role of CSB is given below:

"The future of cloud computing will be permeated with the notion of brokers negotiating relationships between providers of cloud services and the service customers. In this context, a broker might be software, appliances, platforms or suites of technologies that enhance the base services available through the cloud. Enhancement will include managing access to these services, providing greater security or even creating completely new services, "[1].

As the users think about using cloud computing during the emergency situations or during the time of a tight deadline, an important challenge that the face is the identification of the right service provider who can satisfy both their functional and the non functional Quality of Service (QoS) requirements and provision the resources on the fly. Thus they look for a CSB who could handle all the hassles of finding and negotiating the Service Level Agreements (SLA) and monitoring the SLA.

A similar need exists in the 3d animation field, where the 3D animation studios are in need of a CSB who could free them from the hassles of finding the right cloud based render farm service provider. A render farm is nothing but a cluster of computers that complete the task of rendering images in parallel. Since cloud computing has the property of elasticity and enables the scalability of resources that form the cluster of computers, cloud based render farms are being widely used by the 3D studios nowadays.

In this paper, we propose a cloud broker service frame work that enables the selection of the cloud based render farm service provider. A layered architecture of the frame work explains the various components of the CSB frame work. The aggregated Utility function method is used for the cloud based render farm service provider selection.

The rest of the paper is organized as follows: Section 2 briefs about the related works that has been carried out related to this topic. Section 3 describes the proposed layered architecture of the CSB. The Aggregated Utility function methodology used for selection the service provider is explained with an illustrative example in the section 4. The results are discussed in detail in the Section 5. Section 6 concludes the major findings and the future work proposed.

## 2. RELATED WORKS
The concept of service identification and selection is a widely explored concept in the area of web services. Many works has been published related to the web services service selection [2], [3], [4], [5]. Many works related to the cloud service selection and the cloud services broker have been also been published [6], [7], [8], [9].

A notable work in cloud services comparison is done by A. Li, et al. called "CloudCmp" which compares the Quality of services offered by the popular public cloud providers [10], [11], [12]. Ten metrics of the most popular cloud service providers are compared in the "CloudCmp". Examples of the metrics considered are: Elastic computing cluster, persistent storage, intra-cloud and wide area network services. The comparison results enable the cloud service users to predict the performance and cost of deploying their applications on to the cloud even before they do it in real time.

Though many of the works have concentrated on the cloud services selection and provisioning, the render farm service selection is still an unexplored area for research though many 3d studios are using the cloud based render farm in real time scenarios. This work is focused on defining a frame work to facilitate the selection of the cloud based render farms that satisfies the functional and the non functional requirements of the 3d studios. It also facilitates the Service Level Agreements negotiation and monitoring of the services through the partnership with the third party monitoring systems. The third party monitoring systems monitor the services for the specific SLA's and report the violation of the SLA's to the concerned parties. Examples of third party monitoring systems include the Monitis (monitis.com), keynote (keynote.com) and Uptrends (uptrends.com).

## 3. LAYERED ARCHITECTURE OF CSB – AN OVERVIEW

Figure 1 gives the layered architecture of the CSB. The major components of the CSB are the following:

### 3.1 CSB Web Portal

The user provides the requirements of the service to the cloud service provider through the web portal. The CSB validates the user and gets the required functional and the non functional requirements from the end user. Some examples of the functional requirements include the versions of the software that are supported like the 3ds max 2009, Maya 7.0 etc , the render engines supported like the Mental Ray, V-Ray etc, Render node configuration etc. The Non functional requirements include the attributes like the service response time, availability, elasticity etc.

### 3.2 RF Discovery Daemon

It is responsible for discovering the services that match the functional requirements of the end user. The functional requirements of the end user are matched with the render farm offerings and the services that match the requirements are filtered.

### 3.3 The RF Selection Daemon

This daemon is responsible for selecting the right service provider that matches both the functional and the non functional Quality of Service (QoS) requirements of the end user using the aggregated utility function method [14], [15].

### 3.4 SLA Manager

The SLA Manager is responsible for the Service Level Agreement negotiation between the end user and the service providers. It is also responsible for managing the third party SLA monitoring service.

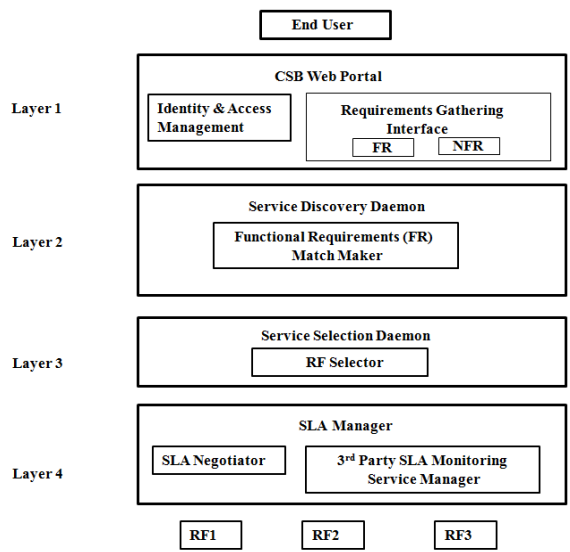

Figure 1: Layered architecture of the CSB.

## 4. THE AGGREGATED UTILITY FUNCTION

The aggregated utility function method is a very simple method that can be used to select a service that needs to satisfy multiple criteria. It lets the user to decide the importance and the sensitivity associated with each attribute by giving the privilege to the end user to specify the weights ($wt_i$) and the sensitivity value ($\beta_i$) assigned to each attribute considered in the selection criteria.

Let Q = { $P_1, P_2, P_3$, -----$P_n$} be the vector of QoS parameters considered in the selection process.

|  | Elasticity | 1/ UT | 1/Cost | 1/SRT | 1/ A |
|---|---|---|---|---|---|
| $Wt_i$ | 0.1 | 0.2 | 0.3 | 0.3 | 0.1 |
| $x_i$ |  | 0.9 | 0.7 | 0.6 | 0.7 |
| pi | 7 | 5 | 9 | 8 | 4 |

**Table 2: Minimum QoS requirements of the end user**

Let Min = {$x_1, x_2, x_3, \text{------} x_n$} with ($0 \leq \text{Min}_i \leq 1$), be the vector of Minimum QoS requirements of the end user for considering the Render Farm in the selection process.

Let $RF_i$ = {$R_1, R_2, R_3, \text{-----} R_k$} be a set of cloud based Render Farm service providers.

Let $QP_i$ = {$q_1, q_2, q_3, \text{-----} q_n$}, with ($0 \leq q_i \leq 1$), be the QoS offerings of the Render Farm service provider $RF_i$.

The Linear aggregate utility function (AU) is defined as follows:

$$AU = wt_1 U_1 = wt_2 U_2 + \text{---} + wt_i U_i \quad (1)$$
with $\sum_1^x wt_i = 1$

$U_i$ – Individual utility function associated with the QoS parameter $P_i$
$Wt_1$ – Weight that the end user assigns to that attribute.

The individual utility function ($U_i$) associated with the QoS parameter $P_i$, takes the sensitivity of the parameter ($\beta_i$) also into consideration. When $\beta_i = 0$, the end user is indifferent to Qos parameter ($P_i$). As the value of ($\beta_i$) increases the sensitivity considered by the end user for the specific parameter also increases. The individual utility function ($U_i$) is calculated using the formula given below which is used in [14] and [15].

$$U_i = P_i^{\beta i} \quad (2)$$

Where,
$\beta_i$ – Measure of the service consumer sensitivity to the QoS Parameter ($P_i$).

## 5. THE SERVICE PROVIDER SELECTION PROCESS

In this section we describe the selection process of the Render Farm service provider using the aggregated utility function method. In order to enable the comparison of the QoS offerings of the various Render farms, the values of the QoS parameters are assumed to be in the normalized form. Thus, the values of the QoS parameters in the normalized form lie between 0 and 1. In case of a QoS attribute with positive tendency, the values are considered as such and a higher normalized value closer to 1 indicates a higher quality. Whereas a lower value closer to 0 indicates a lower quality.

Examples of QoS property with positive tendency are throughput, availability, Elasticity etc. If the attribute has a negative tendency then the inverse of the values are considered. For example: (1/ Cost), (1/ upload time), (1/ServiceResponseTime) etc. The threshold value which is the normalized minimum QoS attribute values that is acceptable by the end user is evaluated using the weight ($wt_i$) and a sensitivity value ($\beta_i$) assigned by the user for all the attributes considered for selection.

An illustrative example is given below to explain the selection process of the Render Farm service provider using the aggregated utility function. The QoS offerings normalized values for the attributes selected by the end user is collected from the potential Render Farm service providers. The Table 1 given below gives the normalized values of the Render farm Service Provider offerings collected from five different Render Farm service providers (RF) for the attributes selected by the end user like the Elasticity, Upload Time (1/UT), Cost, Service Response Time (1/SRT), Availability (1/A). since the attributes like the Upload Time (1/UT), Cost, Service Response Time (1/SRT), Availability (1/A) have negative tendency, which means that the lower values indicate the best quality the inverse value of the attributes are considered in the table. The aggregated utility function value is calculated for each Render Farm service provider using the service provider offerings values given in the Table 1. The minimum QoS attribute value (threshold value) that is acceptable by the end user is calculated using the details of the weight ($wt_i$), sensitivity value ($\beta_i$) and normalized minimum QoS requirement assigned by the user for all the attributes considered for selection as given in the Table 2.

|  | Elasticity | 1/UT | 1/Cost | 1/SRT | 1/A |
|---|---|---|---|---|---|
| RF1 | 0.75 | 0.98 | 0.97 | 0.9 | 0.7 |
| RF2 | 0.8 | 0.96 | 0.8 | 0.95 | 0.85 |
| RF3 | 0.95 | 0.9 | 0.85 | 0.85 | 0.9 |
| RF4 | 0.6 | 0.94 | 0.7 | 0.6 | 0.8 |
| RF5 | 0.85 | 0.92 | 0.8 | 0.75 | 0.9 |

**Table 1: Normalized values of QoS offering of Cloud based render farms**

The aggregated utility function value calculated using the formula (1) and (2) for each Render Farm service provider from the highest to the lowest offer is given below:

$RF_1$ (0.6), $RF_2$ (0.5), $R_3$ (0.4), $RF_5$ (0.3), $RF_4$ (0.2)

Where, the values given between the parentheses correspond to the computed aggregate utility function.

The threshold utility value computed for the minimum quality requirements of the end user is EU (0.230).

## 6. RESULTS AND DISCUSSION

The aggregated utility function value calculated for each Render Farm service providers gives the QoS offerings of the render farm service providers. Since the threshold utility value computed for the minimum quality requirements of the end user is EU (0.230), all the render farms except the $RF_4$ are potential service providers. However the $RF_1$ with the aggregated utility function value as 0.6 is the best choice for the end user followed by $RF_2$ (0.5), $R_3$(0.4), $RF_5$ (0.3).

## 7. CONCLUSION AND FUTURE WORK

The need of a CSB who could free the 3d studios from the hassles of finding the right service providers is very clear and in the future the frame work would be further developed to include the other essential feature of a CSB like the SLA negotiation, SLA monitoring etc which is an important function of the Cloud Broker service. At present the SLA's that are specific to the Render farms are not clearly stated to the users and only a non disclosure agreement is signed between the company and the user. Thus as an extension work we would develop an SLA based framework for the render farm selection. An Ontology is to be created to define the functional requirements of the 3d studios, this would help in the dynamic matching and selection of the render farm services that satisfy the needs of the user.